\documentstyle[12pt,prd,aps,psfig,floats,preprint]{revtex}

\begin{document}
\title{Applying the Inverse Average Magnitude Squared Coherence
Index for Determining Order-Chaos Transition in a System Governed
by H\'enon Mapping Dynamics}

\author{Carlos R. Fadragas\thanks{fadragas@mfc.uclv.edu.cu}}

\address{Physics Department, Faculty of Mathematics, Physics and
Computer Science,\\Central University of Las Villas, Santa Clara,
Cuba}

\author{Rub\'en Orozco Morales\thanks{rorozco@fie.uclv.edu.cu}}

\address{Center for Studying Electronics and Information Technologies,\\
Faculty of Electrical Engineering, Central University of Las
Villas, Santa Clara, Cuba}

\date{\today}
\maketitle \draft

\begin{abstract}
The quantitative determination of the order-chaos transition in a
nonlinear dynamical system described by H\'enon mapping defined as
$x[n+1]=1.0-A*x[n]^2+B*y[n], y[n+1]=B*x[n]$, where $B=0.3$, and
$A$ is an adjustable control parameter, was made. This was
achieved by applying the Inverse Average Magnitude-Squared
Coherence Index (IAMSCI). This method is based on the Welch
average periodogram technique and it has the advantage respect to
nonlinear dynamical methods that it may be applied to any
stationary signal by using discrete Fourier transform (DFT)
representation which allows to operate on a short discrete-time
series. Its effectiveness was demonstrated by comparing the
results obtained by applying IAMSCI method with those obtained by
calculating the largest Lyapounov exponent (LLE), both applied to
the same discrete-time series set derived from the H\'enon
mapping.
\end{abstract}

\newpage
\section{Introduction}
Nonlinear dynamical methods for analyzing a discrete-time series
and for determining order-chaos transition in a dynamical system
have been widely used. But it is recognized by several authors
that nonlinear dynamical methods used imply a high computational
cost. In this sense, therefore, any method that allows to obtain
some information about the behavior of a dynamical system and that
can operate over a relatively shorter discrete-time series,
acquires an important practical value. For complementing those
methods of nonlinear dynamical analysis we shall now consider to
demonstrate that introducing the concept of the inverse average
magnitude-squared coherence index is possible to quantify the
order-chaos transition in a system governed by the H\'enon mapping
dynamics. It is very important when dealing with a chaotic system
keeping in mind that local sensitivity to a small error is the
hallmark of such a system. A dynamical system such those that are
described by the H\'enon mapping dynamics produces a discrete-time
non-chaotic series when the control parameter $A$ value satisfies
the condition $A<1.0532$. For such a condition a discrete-time
series generated by the recursive process does not almost change
if the initial condition is modified and there exists phase
consistency between the two discrete-time series at a given
frequency. However, when the control parameter value is greater
than 1.0532, a discrete-time series generated by the H\'enon
mapping does meaningfully change if the initial condition is
slightly modified and phase consistency between the two
discrete-time series at a given frequency, in general, must be
lost\cite{Fadragas2}. This effect may be quantified by applying an
inverse average magnitude-squared coherence index\cite{Fadragas3}.
The magnitude-squared coherence has been used by several authors
for measuring phase constancy between two or more signals at one
frequency\cite{Thakor,Suresh,WolfE,Eddins,Sih,Liberati} In former
works\cite{Fadragas1,Fadragas2} the DFT representation for
qualitatively describing the order-chaos transition in a dynamical
system is used. By introducing an inverse average
magnitude-squared coherence index (IAMSCI) the order-chaos
transition in a dynamical system described by the logistic
equation was quantitatively determined\cite{Fadragas3}. Now we
intended, following this idea, to make the description of
order-chaos transition in a dynamical system governed by the
H\'enon mapping dynamics by applying the Welch average periodogram
method for calculating the power spectral density and from this,
to introduce the inverse average magnitude-squared coherence index
for indicating quantitatively order-chaos transition in the
dynamical system indicated. The fundamental advantage of this
method upon the others is that it can operate on a relatively
shorter discrete-time series. By estimating the largest Lyapounov
exponent was able to evaluate those results obtained here applying
an inverse average magnitude squared coherence index, defined by
averaging $MSC$ estimate over all the frequency.

\section{Methods and Materials}

\subsection{Obtaining the discrete-time series set}

The H\'enon mapping can be formulated as given by

\begin{eqnarray}
x[n+1]=1.0-A*x[n]^2+B*y[n],\\
y[n+1]=B*x[n]
\end{eqnarray}

where B=0.3, and A is an adjustable control parameter. With the
initial condition (x[1],y[1]) specified, a discrete-time series
with a given length by a recursive action is obtained, each time.
The nature of the behavior of a data sequence obtained can be
modified by chosen conveniently a value of the control parameter
A. The interval [0.2,1.4] was chosen, and a set of evenly spaced
values of A with a step $\delta$=0.01 was considered to obtain a
family consisting of 120 discrete-time series, each of them
containing $N=2^{10}$ samples. The plot of the histogram for each
discrete-time series allows to have immediately a simple
statistical characterization of the discrete-time series. In order
to make easier further computations, data points were organized
into a rectangular matrix containing 120 columns, being each of
them a discrete-time series with the length $N=1024$ data points.

\subsection{Determining the largest Lyapounov exponent $\lambda_1$}

A system containing one or more positive exponents is to be
defined as chaotic. For calculating the Lyapounov exponents
spectrum some algorithms have been developed. One of the most used
algorithm is that reported by Wolf et al\cite{Wolf}. From the
exponents spectrum, the largest exponent, $\lambda_{1}$, decides
the behavior of the dynamical system. For calculating the largest
exponent, one can also refer the algorithm proposed by Rosenstein
et al\cite{Rosenstein}. Details of these algorithms can be found
in refereed papers. The largest Lyapounov exponent,$\lambda_1$,
for each time series in the set was determined using a
professional software\cite{Sprott}. The selection of the time
delay, $\tau$, and the embedding dimension, $D_{e}$, required for
reconstructing the phase space is part of the problem.    The
value $\tau$=1 was chosen considering the optimal filling of the
phase space method\cite{Buzug}. For embedding dimension the value
$D_{e}$=3 was chosen.

\subsection{Estimating the IAMSCI using the Welch average periodogram}

It is highly meaningful when dealing with a chaotic system keeping
in mind that local sensitivity to a small error is the hallmark of
such a system. A dynamical system as the one described by the
logistic equation produces a discrete-time non-chaotic series when
the control parameter $A$ value satisfies the condition
$A<1.0532$. For such a condition a discrete-time series generated
by the recursive process does not almost change if the initial
condition is modified and there exists phase consistency between
the two discrete-time series at a given frequency. However, when
the control parameter value is greater than 1.0532, a
discrete-time series generated by the H\'enon mapping does
meaningfully change if the initial condition is slightly modified
and phase consistency between the two discrete-time series at a
given frequency, in general, must be lost. This effect may be
quantified by applying an inverse average magnitude-squared
coherence index. In order to apply the Welch average periodogram
method for determining the power density spectrum $S_x[k]$ of a
discrete-time series one follows the following steps: (a)
Decompose the sequence of N ($=1024$) data points in M ($=9$)
segments, each of one having the same length L ($=256$), and which
may be overlapped P samples. The usual value $P=0.625*L$ was
chosen. (b) Calculate the DFT representation by fft algorithm for
each segment as given by $X_m[k]=fft(x_m[n])$, where
$m=1,2,\ldots,M$, and $n=0,1,2,\ldots,L-1$. For each segment a
L-fft was calculated. (c) Calculate the periodogram for each
segment $m$ as given by

\begin{eqnarray}
P_m[k]=\frac{1}{L}\vert X_m[k]\vert^2,
\end{eqnarray}

where $m=1,2,\ldots,M$. (d) Finally, calculate the average
periodogram as given by

\begin{eqnarray}
S_x[k]=\frac{1}{M}\sum_{m=1}^M P_m[k],
\end{eqnarray}

where $k=0,1,2,\ldots,L-1$. For a short discrete-time series a
rectangular window is recommended. For determining the
magnitude-squared coherence sequence the procedure described above
is applied to a second signal $y[n]$ obtaining the average
periodogram $S_y[k]$. The average cross-periodogram $S_{xy}[k]$ is
also calculated. The magnitude-squared coherence sequence was
calculated as given by

\begin{eqnarray}
MSC_{xy}[k]=\frac{S_{xy}[k]*S_{xy}^*[k]}{S_{x}[k]*S_{y}[k]}
\end{eqnarray}

The $MSC_{xy}[k]$ sequence describes the phase consistency between
the two signals at a given frequency $k$. The mean value of the
sequence, $\langle MSC_{xy}[k]\rangle$, may be used as an index of
global coherence between the two signals. When dealing with a
chaotic discrete-time series this index must tend toward zero, and
when dealing with a deterministic discrete-time series this index
must tend toward one\cite{Fadragas3}. However, it was found that a
better index may be defined as given by

\begin{eqnarray}
f=10*log_{10}\frac{1}{\langle MSC_{xy}[k]\rangle}.
\end{eqnarray}

where $f$ is given in decibel. It makes easier to compare results
obtained here with those obtained by applying the largest
Lyapounov exponent estimate method.

\section{Results}

Figures, from 1 to 4, show a sample of four representative time
series and their corresponding magnitude-squared coherence
espectra. In order to give a better view of each time series it
was only considered 128 data points of each discrete-time series
in its plotting, but it was completely considered when its
corresponding magnitude-squared coherence was calculated. In
either case, the particular values of $A$ parameter were
indicated. Figure 5 depicts the results of plotting the inverse
average magnitude-squared coherence (1/$\langle
MSC_{xy}[k]\rangle$), calculated for each time series in the set,
as a function of the control parameter $A$. And figure 6 depicts
the results of plotting the inverse average magnitude-squared
coherence index, $f$, in decibels, calculated for each time series
in the set, as a function of the control parameter $A$ too. Figure
7 shows the plotting of the largest Lyapounov exponent estimate,
$\lambda_1$, calculated for each time series in the set, as a
function of the control parameter $A$. This reference serves as a
control plot for the discussion of the results obtained by
applying the inverse average magnitude-squared coherence index
method.

\section{Discussion of the results}

The application of the DFT representation of a chaotic-type signal
gives satisfactory results for determining the order-chaos
transition in a system described by the H\'enon mapping
dynamics\cite{Fadragas2}. A critical value for control parameter
$A$ is reported in literature beyond which a sequence produced by
the H\'enon mapping exhibits a chaotic behavior. This threshold
value approaches $A\cong 1.0532$. On the other hand, figures from
1 to 4 show the time domain representation of some of the time
series and their corresponding magnitude-squared coherence
spectrum. Comparing figures 1 ($A=$) with 2 ($A=$), 3 ($A=$), and
4 ($A=$), it can be deduced the bifurcation effect and the
transition to chaos. Note that in figures 2, 3 and 4 the control
parameter satisfies the condition $A>1.0532$, and it corresponds
to a value for a positive largest Lyapounov exponent, as it can be
estimated from figure 7. It may be observe in those cases that the
magnitude-squared coherence spectrum exhibits values far away from
one for several values of the frequency in the $MSC$ spectrum. It
makes the average magnitude-squared coherence to be less than one,
corresponding to a situation for which the discrete-time series
becomes incoherent indicating in this case a chaotic behavior of
the dynamical system. This situation can be observed in figures 5
and 6 where either the quantity (1/$\langle MSC_{xy}[k]\rangle$)
as the inverse average $MSC$ index, $f$, are greater than $1$ and
$0$, respectively, for $A>1.0532$. Applying the inverse average
magnitude-squared coherence index (IAMSCI) to a discrete-time
series in a family derived from an observable in a dynamical
system is recommended for quantitatively detecting the order-chaos
transition, and this can be added to the metric tools of the
nonlinear dynamical analysis for complementing the research of a
discrete-time series.

\section{Conclusions}

The application of the inverse average magnitude-squared coherence
index (IAMSCI), $f$, to a discrete-time series obtained from the
H\'enon mapping can be done on a chaotic discrete-time series.
This discrete-time series has the property that exhibits a high
level of coherence lost, and it can be detected by applying the
IAMSCI method. This method does not exert a strong requirement on
the length of the experimental data sequence. If one deals with a
nonlinear dynamical system which can modify its behavior between
order and chaos, a discrete-time series produced by that system
reflects this change, and the IAMSCI method applying to the
discrete-time series allow to determine this change. Any method
that permits some quantitative evaluation about the behavior of a
dynamical system and that can work over a relatively short time
series acquires a particular importance.

\section*{Acknowledgement}

We acknowledge Pedro A. S\'anchez Fern\'andez from the Department
of Foreign Languages for the revision of the English version, and
Yoelsy Leiva for the edition. We also thank the Ministry of Higher
Education of Cuba for partially financial support of the research.

\newpage
\section{Figures}

\newpage
\begin{figure}[tbh!]
\centerline{\psfig{figure=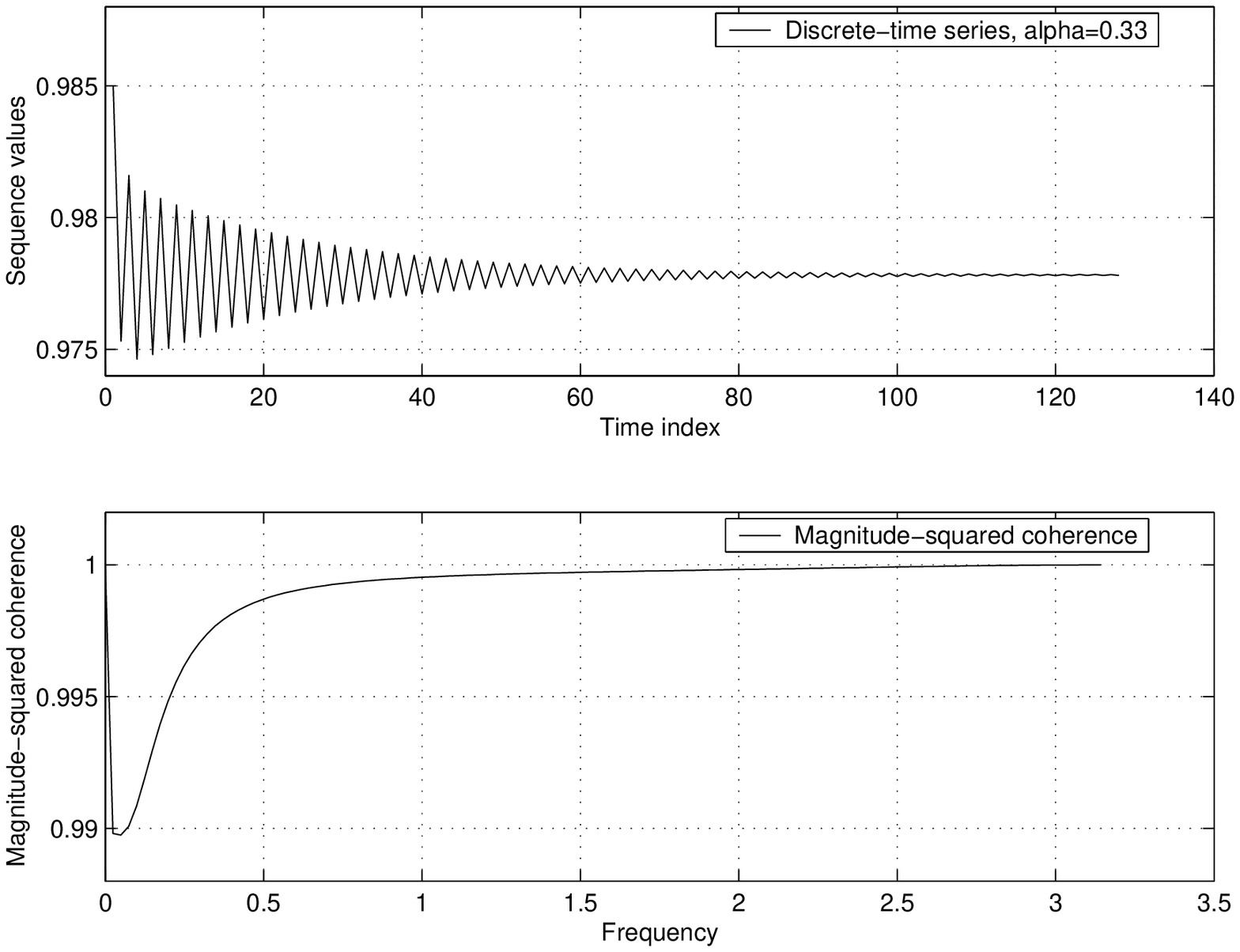,width=0.5\textwidth,angle=0}}
\bigskip
\caption{Discrete-time series and its magnitude-squared coherence,
for A=0.33} \label{fig1}
\end{figure}

\begin{figure}[tbh!]
\centerline{\psfig{figure=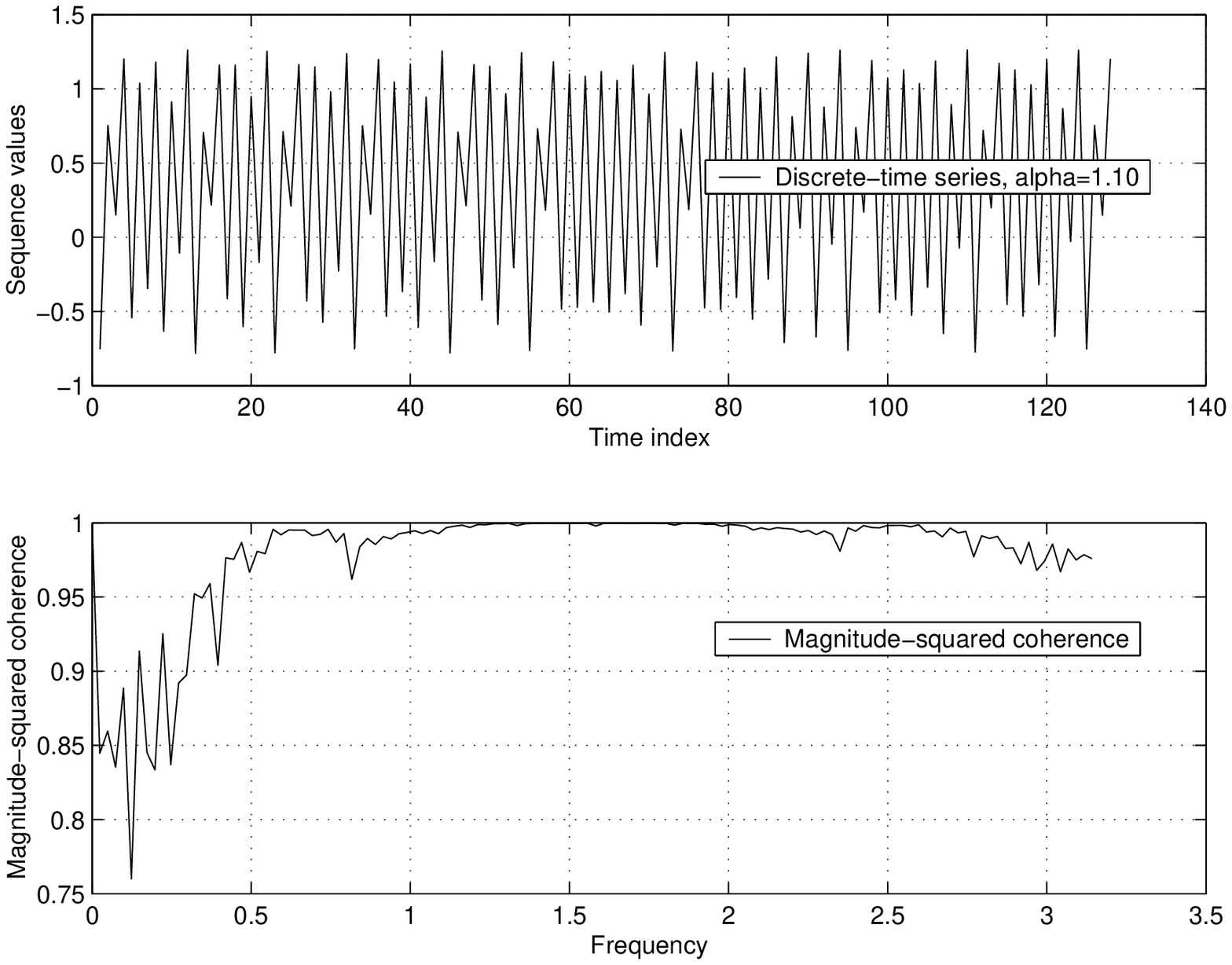,width=0.5\textwidth,angle=0}}
\bigskip
\caption{Discrete-time series and its magnitude-squared coherence,
for A=1.10} \label{fig2}
\end{figure}

\begin{figure}[tbh!]
\centerline{\psfig{figure=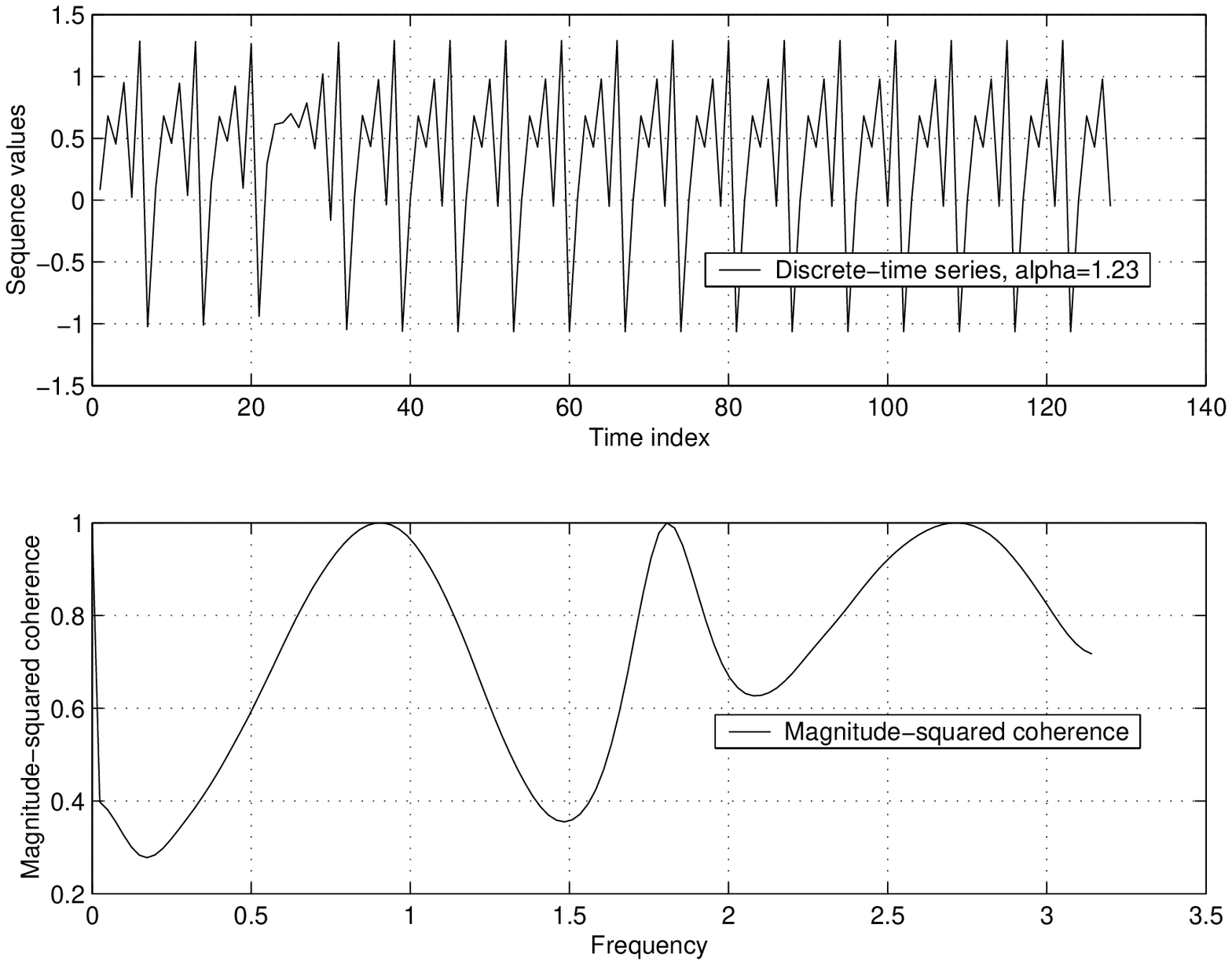,width=0.5\textwidth,angle=0}}
\bigskip
\caption{Discrete-time series and its magnitude-squared coherence,
for A=1.23} \label{fig3}
\end{figure}

\begin{figure}[tbh!]
\centerline{\psfig{figure=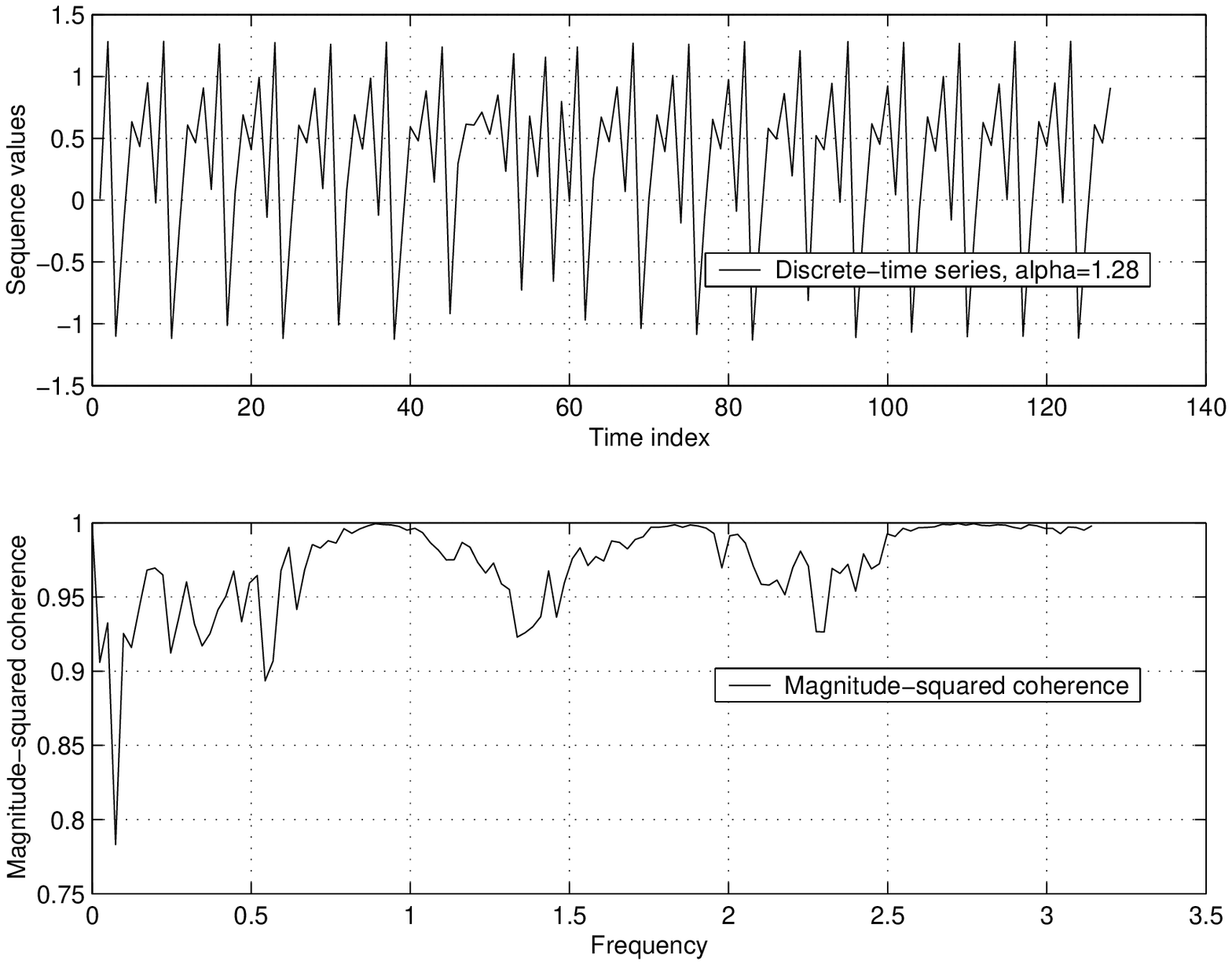,width=0.5\textwidth,angle=0}}
\bigskip
\caption{Discrete-time series and its magnitude-squared coherence,
for A=1.28} \label{fig4}
\end{figure}

\begin{figure}[tbh!]
\centerline{\psfig{figure=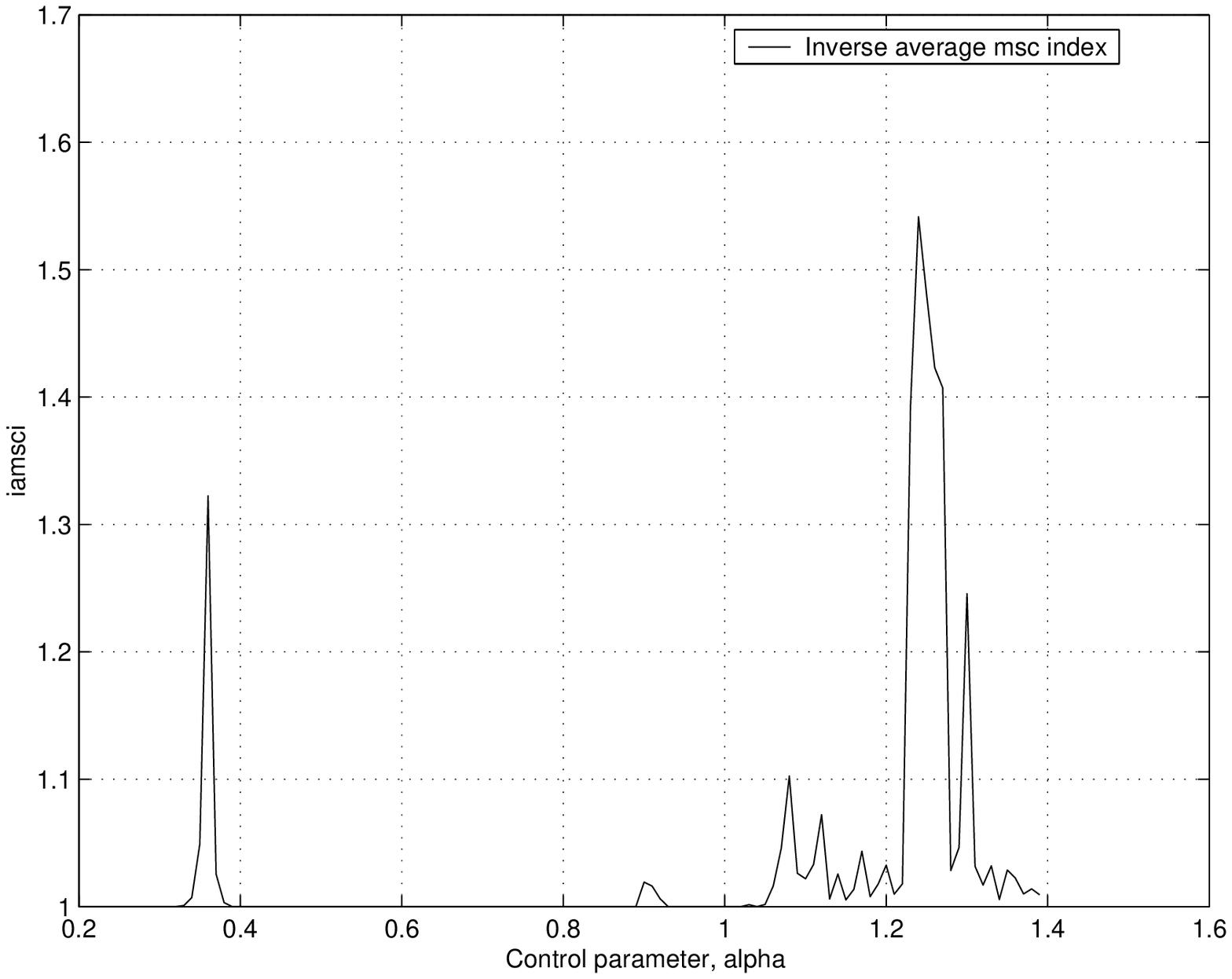,width=0.5\textwidth,angle=0}}
\bigskip
\caption{Inverse average magnitude-squared coherence index}
\label{fig5}
\end{figure}

\begin{figure}[tbh!]
\centerline{\psfig{figure=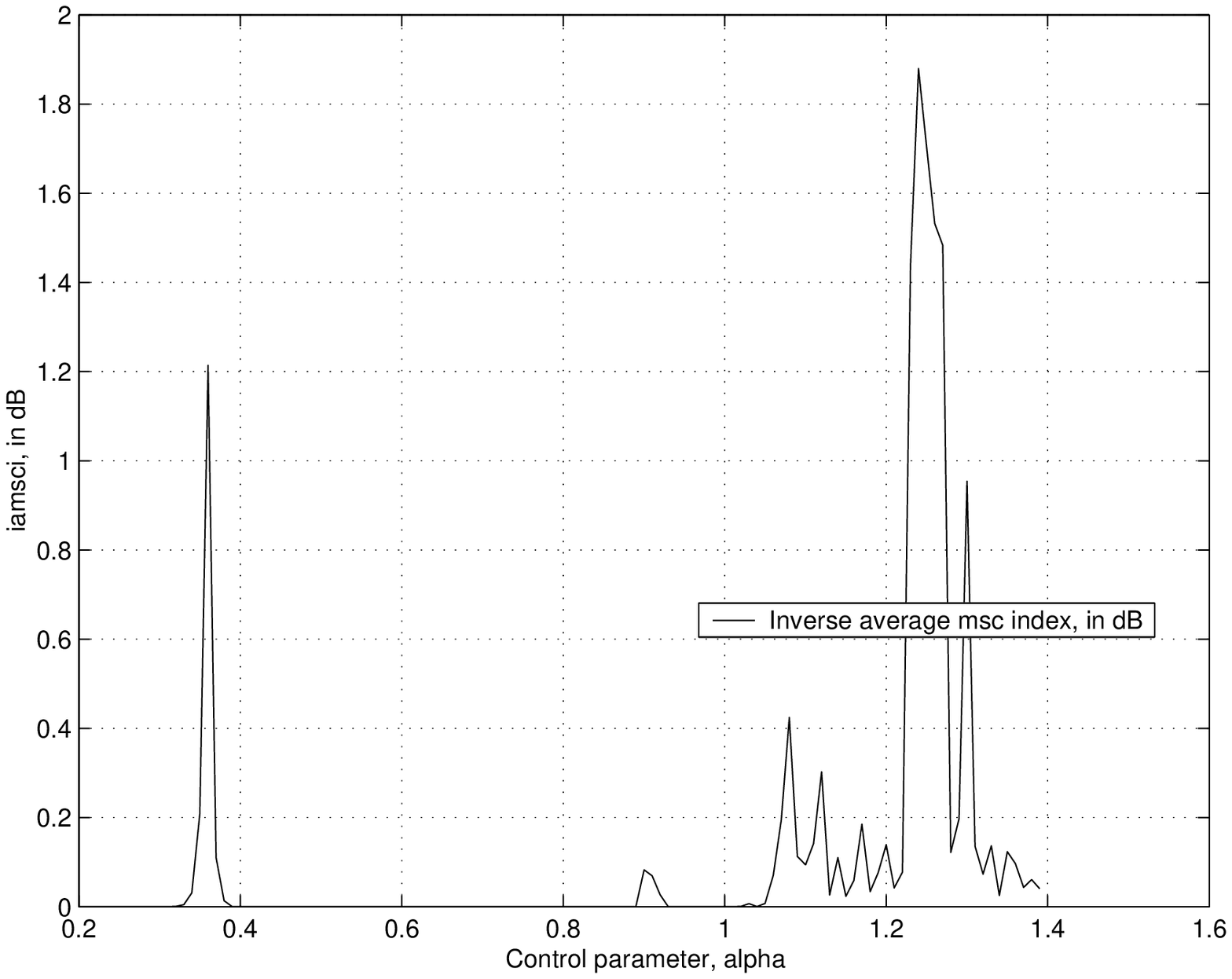,width=0.5\textwidth,angle=0}}
\bigskip
\caption{Inverse average magnitude-squared coherence index, in dB}
\label{fig6}
\end{figure}

\begin{figure}[tbh!]
\centerline{\psfig{figure=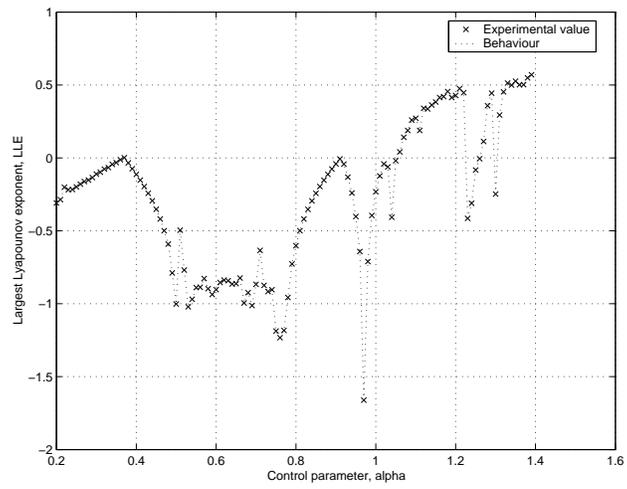,width=0.5\textwidth,angle=0}}
\bigskip
\caption{Largest Lyapounov versus control parameter r}
\label{fig7}
\end{figure}

\end{document}